\documentclass[lettersize,journal]{IEEEtran}

\usepackage{amsmath}
\usepackage{array}
\usepackage{multicol}
\usepackage{multirow}
\usepackage{hhline} 
\usepackage{amssymb}
\usepackage{mathtools}
\usepackage{lipsum}
\usepackage{color, soul}
\usepackage{xcolor}
\usepackage{framed}
\usepackage{times, epsfig}
\usepackage{amscd}
\usepackage{graphicx}
\usepackage[tight,footnotesize]{subfigure}
\usepackage{epstopdf}
\usepackage{pdfpages}
\usepackage{tablefootnote}
\usepackage{tabularx, booktabs}
\usepackage{cancel}
\usepackage[normalem]{ulem}
\usepackage{cuted}
\usepackage[acronym]{glossaries}
\usepackage{algorithmic} 
\usepackage[linesnumbered,ruled]{algorithm2e}
\SetKwRepeat{Do}{do}{while}
\usepackage[utf8]{inputenc}
\usepackage{textcomp}

\usepackage{graphicx}
\usepackage{caption}
\usepackage{subcaption}
\usepackage{float}

\def\approxprop{%
  \def\p{%
    \setbox0=\vbox{\hbox{$\propto$}}%
    \ht0=0.6ex \box0 }%
  \def\s{%
    \vbox{\hbox{$\sim$}}%
  }%
  \mathrel{\raisebox{0.7ex}{%
      \mbox{$\underset{\s}{\p}$}%
    }}%
}

\newcommand{\todo}[1]{{\leavevmode\color{magenta}{#1}}}

\def\be{ \begin{eqnarray} }
\def\ee{ \end{eqnarray} }
\def\bit { \begin{item} }
\def\eit { \end{item} }
\def\bnum { \begin{enumerate} }
\def\enum { \end{enumerate} }

\makeatletter
\let\save@mathaccent\mathaccent
\newcommand*\if@single[3]{%
  \setbox0\hbox{${\mathaccent"0362{#1}}^H$}%
  \setbox2\hbox{${\mathaccent"0362{\kern0pt#1}}^H$}%
  \ifdim\ht0=\ht2 #3\else #2\fi
  }
\newcommand*\rel@kern[1]{\kern#1\dimexpr\macc@kerna}
\newcommand*\wideaccent[2]{\@ifnextchar^{{\wide@accent{#1}{#2}{0}}}{\wide@accent{#1}{#2}{1}}}
\newcommand*\wide@accent[3]{\if@single{#2}{\wide@accent@{#1}{#2}{#3}{1}}{\wide@accent@{#1}{#2}{#3}{2}}}
\newcommand*\wide@accent@[4]{%
  \begingroup
  \def\mathaccent##1##2{%
    \let\mathaccent\save@mathaccent
    \if#42 \let\macc@nucleus\first@char \fi
    \setbox\z@\hbox{$\macc@style{\macc@nucleus}_{}$}%
    \setbox\tw@\hbox{$\macc@style{\macc@nucleus}{}_{}$}%
    \dimen@\wd\tw@
    \advance\dimen@-\wd\z@
    \divide\dimen@ 3
    \@tempdima\wd\tw@
    \advance\@tempdima-\scriptspace
    \divide\@tempdima 10
    \advance\dimen@-\@tempdima
    \ifdim\dimen@>\z@ \dimen@0pt\fi
    \rel@kern{0.6}\kern-\dimen@
    \if#41
      #1{\rel@kern{-0.6}\kern\dimen@\macc@nucleus\rel@kern{0.4}\kern\dimen@}%
      \advance\dimen@0.4\dimexpr\macc@kerna
      \let\final@kern#3%
      \ifdim\dimen@<\z@ \let\final@kern1\fi
      \if\final@kern1 \kern-\dimen@\fi
    \else
      #1{\rel@kern{-0.6}\kern\dimen@#2}%
    \fi
  }%
  \macc@depth\@ne
  \let\math@bgroup\@empty \let\math@egroup\macc@set@skewchar
  \mathsurround\z@ \frozen@everymath{\mathgroup\macc@group\relax}%
  \macc@set@skewchar\relax
  \let\mathaccentV\macc@nested@a
  \if#41
    \macc@nested@a\relax111{#2}%
  \else
    \def\gobble@till@marker##1\endmarker{}%
    \futurelet\first@char\gobble@till@marker#2\endmarker
    \ifcat\noexpand\first@char A\else
      \def\first@char{}%
    \fi
    \macc@nested@a\relax111{\first@char}%
  \fi
  \endgroup
}
\makeatother

\newcommand\doubleoverline[1]{\overline{\overline{#1}}}
\newcommand\widebar{\wideaccent\overline}
\newcommand\widebarbar{\wideaccent\doubleoverline}

\makeglossaries

\newacronym{RMSProp}{RMSProp}{root mean square propagation}
\newacronym{GDM}{GDM}{gradient descent with momentum}
\newacronym{BARProp}{BARProp}{buffer-aided RMSProp}
\newacronym{SGD}{SGD}{stochastic gradient descen}
\newacronym{RSS}{RSS}{received signal strength}
\newacronym{TN}{TN}{target node}
\newacronym{WSN}{WSN}{wireless sensor network}
\newacronym{GPS}{GPS}{global positioning system}
\newacronym{AN}{AN}{anchor node}
\newacronym{RMSE}{RMSE}{root mean square error}
\newacronym{SOCP}{SOCP}{second-order cone programming}
\newacronym{SDP}{SDP}{semidefinite programming}
\newacronym{LSRE}{LSRE}{least squared relative error}
\newacronym{TOA}{TOA}{time of arrival}
\newacronym{AOA}{AOA}{angle of arrival}
\newacronym{NLS}{NLS}{nonlinear least squares}
\newacronym{UAV}{UAV}{unmanned aerial vehicle}
\newacronym{FIFO}{FIFO}{first-in-first-out}

\begin{document}

\title{\huge BARProp: Fast-Converging and Memory-Efficient RSS-Based Localization Algorithm for IoT}


\author{ Luis F. Abanto-Leon, and 
	Muhammad Salman, and
    Lismer Andres Caceres-Najarro 
    \thanks{ This work was supported in part
    by the National Research Foundation of Korea under Grant NRF-2021R1I1A1A01041257 and in
    part by 
    by the research fund from Chosun University, 2024.
    (Corresponding author: Lismer Andres Caceres
    Najarro)}
	\thanks{L. F. Abanto-Leon is with the Faculty of Electrical Engineering and Information Technology (ETIT), Ruhr-Universität Bochum, Germany. Email: l.f.abanto@ieee.org.}
     \thanks{L. A. Caceres-Najarro are with the Department of Computer Science of Chosun University,
       Gwangju 61452,
     Republic of Korea. Email: andrescn@chosun.ac.kr}
        \thanks{M. Salman is affiliated with the Faculty of Computer Science and Engineering at the Ghulam Ishaq Khan Institute of Engineering Sciences and Technology, Topi, Sawabi, Pakistan. Email: salman@nsl.inha.ac.kr}
}


\maketitle

\setlength{\baselineskip}{1\baselineskip}

\begin{abstract}



Leveraging \gls{RSS} measurements for indoor localization is highly attractive due to their inherent availability in ubiquitous wireless protocols. However, prevailing \gls{RSS}-based methods often depend on complex computational algorithms or specialized hardware, rendering them impractical for low-cost access points. To address these challenges, this paper introduces \gls{BARProp}, a fast and memory-efficient localization algorithm specifically designed for \gls{RSS}-based tasks. The key innovation of \gls{BARProp} lies in a novel mechanism that dynamically adapts the decay factor by monitoring the energy variations of recent gradients stored in a buffer, thereby achieving both accelerated convergence and enhanced stability. Furthermore, \gls{BARProp} requires less than $15\%$ of the memory used by state-of-the-art methods. Extensive evaluations with real-world data demonstrate that \gls{BARProp} not only achieves higher localization accuracy but also delivers at least a fourfold improvement in convergence speed compared to existing benchmarks.

\end{abstract}

\begin{IEEEkeywords}
Localization, 
adaptive algorithms, 
gradient descent, 
learning,
received signal strength.
\end{IEEEkeywords}

\IEEEpeerreviewmaketitle

\glsresetall

\section{Introduction}\label{sec_Intro}

The internet of things (IoT) has expanded rapidly, with connected devices now surpassing 30 billion globally \cite{lombardi2021internet}. The sensing backbone of IoT systems is increasingly shifting to wireless technology because it offers flexibility, scalability, and cost efficiency that wired setups cannot match. The sensing side of IoT is comprised of \glspl{WSN} which seamlessly collect real-world data from distributed environments, enabling efficient and adaptable IoT deployments \cite{gulati2022review, yuan2025wsan}.

\Glspl{WSN} play a crucial role in numerous applications such as smart housing systems that optimize energy consumption and enhance residential comfort, smart cities that integrate various urban services for improved quality of life, autonomous navigation systems that enable self-driving vehicles and robotic mobility, precision smart farming techniques that maximize agricultural productivity while minimizing resource waste, and advanced robotics applications that require real-time environmental awareness \cite{caceres2022fundamental}. These applications rely on the accurate monitoring and control capabilities provided by \glspl{WSN}, which consist of spatially distributed sensor nodes that collect and transmit data. With advancements in wireless technologies such as 5G and 6G that promise even more transformative capabilities, including ubiquitous connectivity, artificial intelligence integration, and holographic communications,
the efficiency and operational capabilities of individual sensor nodes are expected to increase dramatically \cite{chowdhury2020_6G}.


These technological advancements will empower sensor nodes to process complex data locally, enable ultra-fast and reliable communication, and support advanced sensing modalities \cite{jamshed2022challenges}. As a result, these advancements will further embed \glspl{WSN} as indispensable components of modern technological infrastructures. However, the full potential of these \glspl{WSN} can only be realized when precise knowledge of sensor node locations \cite{yadav2023systematic}\footnote{This location-data correlation is critical because environmental measurements such as temperature readings, humidity levels, motion detection, or chemical concentrations are meaningful only when their precise spatial origins are known. Without accurate localization, even the most sophisticated sensor data becomes contextually ambiguous and potentially misleading.}. As the information retrieved by these nodes is intrinsically linked to their locations, accurate sensor node localization is paramount for the effectiveness of \gls{WSN} applications. Despite the critical importance of accurate localization, traditional positioning solutions present significant challenges when applied to \gls{WSN} environments.


While \gls{GPS} technology has proven highly capable of providing precise localization with accuracy typically within a few meters under optimal conditions, it is often impractical for \glspl{WSN} due to its high energy consumption, significant costs, and the requirement for line-of-sight conditions. 
These limitations have spurred the development of various \gls{GPS}-free localization techniques, which utilize alternative measurements such as \gls{RSS}, \gls{TOA}, and \gls{AOA} \cite{fomichev2022next2you}. Among these, \gls{RSS}-based localization is particularly popular due to its accessibility, minimal hardware requirements, lower power consumption, and suitability across diverse environments. \Gls{RSS} measurements, obtained from the attenuation of signal strength with distance, provide a viable means for estimating node locations without the need for specialized infrastructure.

Recent advancements in localization have prioritized improving accuracy through various convex optimization methodologies. Traditional approaches to node localization commonly rely on \gls{NLS} formulations, which are typically addressed inexactly via \gls{SOCP} and \gls{SDP} relaxations \cite{jour_Tomic,jour_Chang_2018,jour_Z_Wang_2019,wang2022SDP_biased_RSS,mukhopadhyay2022invex,lu2024racln,jiang2024cooperative}. For instance, the work in \cite{wang2022SDP_biased_RSS} considered RSS-based localization and formulated the problem as a relaxed SDP to enable tractable solutions. Additionally, \cite{mukhopadhyay2022invex} approximated the localization cost function by an invariant convex formulation, which was then solved using a gradient descent algorithm. While this method offers lower computational complexity than \cite{wang2022SDP_biased_RSS}, it also yields reduced localization accuracy. Among the existing optimization-based methods, \cite{jour_Z_Wang_2019} achieves the best localization performance, albeit at the cost of substantially higher computational complexity. In general, although SDP- and SOCP-based algorithms can provide reasonable localization accuracy, their intensive computational requirements render them less suitable for IoT deployments, where sensor nodes are typically constrained in terms of both processing capability and energy resources.

In parallel, machine learning and fingerprinting have emerged as two prominent techniques for location estimation. For instance, a recent approach combined deep convolutional neural networks (CNNs) with continuous wavelet transform (CWT) to estimate \gls{TN} positions \cite{elsisi2024robust}. Here, CWT is applied to convert one-dimensional RSS measurements into two-dimensional feature representations, which are subsequently utilized to train the CNN model using the labeled position data. To mitigate the high cost of site surveys required for RSS-based fingerprinting, a semi-supervised deep learning framework was introduced in \cite{xiang2025modefa}, where fingerprint data are collected at randomly chosen locations and enhanced using a recorrupt-to-recorrupt denoising strategy. Similarly, \cite{dai2024gridloc} proposed a deep learning model that integrates both \gls{RSS} and channel state information to further improve localization accuracy. While these methods demonstrate promising results, they remain inherently scenario-dependent and typically require extensive offline measurements to construct and maintain reliable fingerprint maps.

To address the computational challenges of traditional convex optimization-based methods, evolutionary algorithms have gained considerable attention due to their ability to deliver comparable localization accuracy at reduced computational complexity. For instance, an updated version of the elephant herding optimization algorithm has been shown to outperform \gls{SOCP}-based methods in acoustic localization scenarios, providing a more efficient solution \cite{correia2020MetaHeuristic}. Similarly, an enhanced differential evolution algorithm has demonstrated superior performance in \gls{RSS}-based localization compared to \gls{SOCP} and \gls{SDP} solutions \cite{caceres2020DEOR,caceres2020fastDEOR}. However, the effectiveness of these evolutionary algorithms is often highly dependent on the careful tuning of parameters, which can limit their adaptability and generalizability.

More recently, optimization techniques in big data analytics and deep neural networks have advanced beyond classical \gls{SGD}, giving rise to methods such as \gls{GDM} \cite{qian1999:gdm} and \gls{RMSProp} \cite{tieleman2012:rmsprop}. Both approaches aim to address the fluctuating gradients and unstable updates of \gls{SGD}, which often lead to slow convergence and poor local optima. Specifically, \gls{GDM} enhances stability by incorporating momentum from previous iterations, while \gls{RMSProp} adaptively scales the learning rate using exponentially weighted averages of past squared gradients. Owing to their efficiency, these approaches have been applied to various localization problems. For example, \gls{GDM} has been utilized for cooperative positioning among \glspl{UAV} in outdoor environments \cite{han2024secure}, while \gls{RMSProp} has facilitated indoor localization through differentiable ray tracing techniques \cite{han2025raylocwirelessindoorlocalization}. Nevertheless, both \gls{SGD} and \gls{RMSProp} continue to exhibit limitations when applied to highly non-convex functions, an inherent challenge in localization tasks.
 
Motivated by these observations, this paper addresses the localization problem based on \gls{RSS} measurements and introduces a new optimization algorithm, termed \gls{BARProp}.The proposed approach achieves higher localization accuracy and faster convergence while requiring significantly less memory than state-of-the-art methods. The main contributions of this work are summarized as follows:
\begin{itemize}
	\item \gls{BARProp} builds upon \gls{RMSProp} and incorporates an adaptive decaying factor that dynamically adjusts exponential weights. This is implemented through a \gls{FIFO} buffer that stores the most recent gradients. The design enables large exploratory steps during the early iterations, accelerating convergence, while gradually reducing step sizes over time without diminishing the learning rate to prevent stagnation.
	
	\item To enhance localization accuracy, \gls{BARProp} employs an initialization phase that selects a feasible starting point. Candidate positions are randomly generated, and the most promising one is chosen to initialize the optimization, effectively serving as a stochastic sampling mechanism that trades a small overhead for faster convergence. In addition, to avoid divergence, particularly during the early iterations when step sizes are large, \gls{BARProp} introduces a bounding phase. Estimated positions that fall outside the predefined search area are automatically re-mapped inside the region via controlled random steps.
	
	\item Extensive simulations with real-world data demonstrate that \gls{BARProp} significantly outperforms state-of-the-art localization algorithms in both accuracy and computational efficiency. In particular, it achieves superior localization accuracy compared to all benchmarks while providing a speedup of at least $4$ times relative to the closest competing method. Moreover, although not the primary focus, \gls{BARProp} requires less than \todo{$15\%$} of the memory needed by competing methods.
\end{itemize}


\emph{Notation}: Matrices and vectors are denoted by $ \mathbf{A} $ and $ \mathbf{a} $, respectively. The transpose of $ \mathbf{A} $ is denoted by $ \mathbf{A}^\mathrm{T} $, and the element in the $ r $-th row and $ c $-th column is written as $ \left[ \mathbf{A} \right]_{r,c} $. The space of all $ P \times Q $ real matrices is denoted by $ \mathbb{R}^{P \times Q} $, while $ \left\|  \cdot \right\| _2 $ represents the $ \ell_2 $-norm. The Gaussian and uniform distributions with mean $ \mu $ and variance $ \xi^2 $ are denoted by $ \mathcal{N} \left( \mu, \xi^2 \right) $ and $ \mathcal{U} \left( \mu, \xi^2 \right) $, respectively. In addition, $ \ln \left( \cdot \right) $ denotes the natural logarithm, $ \lfloor \cdot \rfloor $ the floor operator, and $ \mathrm{max} \left\lbrace \cdot \right\rbrace $ and $ \mathrm{min} \left\lbrace \cdot \right\rbrace $ the maximum and minimum operators. Symbols $ \odot $ and $ \oslash $ denote product- and division-wise operations.

\section{System Model} \label{sec_Model}


We assume a two-dimensional \gls{WSN} for simplicity, though the extension to three dimensions is straightforward. 
Consider a \gls{TN} with unknown position $\mathbf{x} = {\left[ x_1, x_2 \right]^\mathrm{T}}$, where $ \mathbf{x} \in \mathbb{R}^2 $. The \gls{WSN} consists of $ N $ \glspl{AN} with known positions, deployed within a region with limits $ X_1^\mathrm{min} $, $ X_1^\mathrm{max} $, $ X_2^\mathrm{min} $, and $ X_2^\mathrm{max} $, such that $  x_1 \in \left[ X_1^\mathrm{min}, X_1^\mathrm{max} \right]  $ and $ x_2 \in \left[ X_2^\mathrm{min}, X_2^\mathrm{max} \right] $. The position of the $n$-th \gls{AN} is denoted by $\mathbf{s}_n \in \mathbb{R}^2$, for $ n = 1, 2, \dots,N $.
Using the path-loss model in \cite{book_rappaport_wireless_comm_1996}, the \gls{RSS} measurement (in dBm) at the $n$-th \gls{AN} is 
\begin{equation} \label {eq_PL}
	P_n = P_0 - 10\gamma \log_{10}\left\| \mathbf{x} - \mathbf{s}_n \right\|_2 + v_n,
\end{equation}
where $ \left\| \mathbf{x} - \mathbf{s}_n \right\|_2 $ represents the distance between $ \mathbf{x} $ and $ \mathbf{s}_n $. Here, $P_0$ represents the \gls{TN}'s transmit power, $\gamma$ denotes the path-loss exponent, and ${v_n}\sim\mathcal{N}(0,\sigma _n^2)$ represents log-shadowing, which is modeled as additive noise.

The maximum likelihood estimator of the \gls{TN}'s position is formulated as
\begin{equation*} \label {eq_ML}
	\mathcal{P}: \hat{\mathbf{x}} = \mathop {\arg \min } \limits_{\mathbf{x}} \sum\limits_{n = 1}^N \frac{ \left( P_n - P_0 + 10\gamma \log_{10} \left\|\mathbf{x}- \mathbf{s}_n \right\|_2 \right)^2}{\sigma_n^2}. 
\end{equation*}

\section{Proposed Localization Algorithm}\label{sec_Approach}

Problem $ \mathcal{P} $ is highly non-convex, and obtaining a globally optimal solution is generally infeasible. Even in cases where such solutions exist, they cannot be computed in real time to support the stringent latency requirements of \gls{WSN} and IoT applications. Furthermore, due to the presence of numerous local optima, traditional convex optimization-based methods can easily become trapped in suboptimal solutions, resulting in degraded localization performance. To overcome these challenges, we propose the \gls{BARProp} algorithm, an enhanced variant of the well-known \gls{RMSProp} method \cite{tieleman2012:rmsprop}, incorporating modifications that accelerate convergence and improve accuracy in \gls{RSS}-based localization.

\textbf{Preliminaries:} \gls{RMSProp} is an adaptive learning rate optimization algorithm that addresses the diminishing step sizes often encountered with \gls{SGD}. It computes an exponentially weighted moving average of the squared gradients. This mechanism helps to stabilize learning by preventing extreme fluctuations in the weight updates. However, a significant limitation of \gls{RMSProp} is that the influence of earlier, potentially suboptimal, gradients can persist in the weighted average, hindering convergence.

To mitigate this issue, \gls{BARProp} introduces a dynamic adjustment of the decay factor. This is achieved by analyzing the variations of squared gradients stored in a short-term buffer. This adaptive mechanism allows \gls{BARProp} to detect and respond to gradual changes in the gradient landscape, reducing the influence of outdated gradients via a dynamically adjusted decay factor. This, in turn, accelerates convergence and enhances the overall performance of the optimization process. The \gls{BARProp} algorithm is detailed in the following.

\textbf{Initial feasible point:} Within the localization region of interest, we generate $ U $ random positions which are stored in set $ \mathcal{Y} = \left\lbrace \mathbf{y}_1, \dots, \mathbf{y}_U \right\rbrace $. From this set, we choose an initial feasible position $ \widetilde{\mathbf{x}}_0 $, as shown below
\begin{align} \label{eq:intial-feasible-point}
	\widetilde{\mathbf{x}}_0 = \mathop{\arg \min}_{ \mathbf{y}_u \in \mathcal{Y} } f \left( \mathbf{y}_u \right), 
\end{align} 
where $ f \left ( \mathbf{x} \right ) = \textstyle \sum\limits_{n = 1}^N \frac{1}{\sigma _n^2} \left( P_n - P_0 + 10\gamma \log_{10} \left\|\mathbf{x}- \mathbf{s}_n \right\|_2 \right)^2 $.
Here, $ \widetilde{\mathbf{x}}_0 $ serves as the initial point for beginning the iterative process to estimate the unknown position of the \gls{TN}.

\textbf{Gradient:} Let $ \widetilde{\mathbf{x}}_j $ be the estimated \gls{TN}'s position at the $ j $-th iteration. Additionally, let $ g \left( \widetilde{\mathbf{x}}_j \right) \in \mathbb{R}^{2} $ be the gradient of $ f \left( \mathbf{x} \right) $ with respect to $ \mathbf{x} $ evaluated at $ \widetilde{\mathbf{x}}_j $, which is given by
\begin{align} \label{eq:gradient}
	\widebar{g} \left( \widetilde{\mathbf{x}}_j \right) = \tfrac{20 \gamma}{ \ln 10 } \textstyle \sum_{n = 1}^N \frac{ h_n \left( \widetilde{\mathbf{x}}_j \right)^2 }{ \sigma_n^2 \left\| \widetilde{\mathbf{x}}_j - \mathbf{s}_n \right\|_2^2 } \left( \widetilde{\mathbf{x}}_j- \mathbf{s}_n \right),
\end{align}
where $ h_n \left( \widetilde{\mathbf{x}}_j \right) = P_n - P_0 + 10\gamma \log_{10} \left\| \widetilde{\mathbf{x}}_j - \mathbf{s}_n \right\|_2 $.

\textbf{Position estimation:} The accuracy of the \gls{TN}'s location is iteratively refined by taking steps in the direction opposite to the gradient, as shown below
\begin{align} \label{eq:position-update}
	\widetilde{\mathbf{x}}_j = \widetilde{\mathbf{x}}_{j-1} - \mu \cdot g \left( \widetilde{\mathbf{x}}_j \right) \oslash \left( \boldsymbol{\delta} + \sqrt{\mathbf{c}_j} \right),
\end{align}
where $ \mu \in \mathbb{R} $ is the learning rate, $ \boldsymbol{\delta} \in \mathbb{R}^{2 } $ is the numerical stability constant, and $ \mathbf{c}_j  $ is a damping term at iteration $ j $ that stabilizes learning and whose computation is detailed next.

\textbf{Smoothed squared gradients:} The squared gradient at iteration $ j $ is computed as $ \widebarbar{g} \left( \widetilde{\mathbf{x}}_j \right) = \widebar{g} \left( \widetilde{\mathbf{x}}_j \right) \odot \widebar{g} \left( \widetilde{\mathbf{x}}_j \right) $, which is used to obtain a smoother version of itself, given by
\begin{align} \label{eq:smoothed-term}
	\mathbf{c}_j = \boldsymbol{\rho}_j \odot \mathbf{c}_{j-1} + \left( \mathbf{1} - \boldsymbol{\rho}_j \right) \odot \widebarbar{g} \left( \widetilde{\mathbf{x}}_j \right),
\end{align}
which is an exponentially smoothed average of the past squared gradients. In addition, $\boldsymbol{\rho}_j$ is the decay factor at iteration $j$, which remains constant in \gls{RMSProp}, whereas \gls{BARProp} dynamically adjusts this parameter based on the gradient behavior, as detailed next.

\textbf{Adaptive decay factor:} We first implement a \gls{FIFO} buffer to retain critical recent information that will be exploited to adapt the decay factor. Thus, we define the buffer index $ \ell $ as
\begin{align} \label{eq:buffer-indexing}
	\ell = j - L\left\lfloor{\tfrac{j-1}{L}}\right\rfloor, 
\end{align}
where $ j $ is the iteration index. The buffer is represented by matrix $ \mathbf{B} \in \mathbb{R}^{2 \times L} $, where its dimensions correspond to the two coordinates, $ x_1 $ and $ x_2 $, and its length $ L $. In particular, the buffer stores the most recent $ L $ squared gradients, which are used to  adjust the decay factor in response to energy level changes. At iteration $ j $, the two elements of the squared gradient are stored in the buffer, as follows
\begin{align} \label{eq:squared-gradients-storing}
	\left[ \mathbf{B} \right]_{k,\ell} = \left[ \widebarbar{g} \left( \widetilde{\mathbf{x}}_j \right) \right]_k,
\end{align} 
where $ k = \left\lbrace 1, 2 \right\rbrace $ is used to refer to each of the coordinates. The adaptive decay factor at iteration $ j $ is computed as
\begin{align} \label{eq:decay-factor}
	\left[ \boldsymbol{\rho}_j \right]_k = \max \left\lbrace \left[ \widetilde{\boldsymbol{\rho}} \right]_k, \left[ \boldsymbol{\gamma}_j \right]_k \right\rbrace  
\end{align} 
where $ \widetilde{\boldsymbol{\rho}} \in \mathbb{R}^{2 } $ is the nominal decay factor that acts as a limit, while $ \boldsymbol{\gamma}_j $ tracks the energy fluctuations of the gradients stored in the buffer, which is defined as follows
\begin{align} \label{eq:gamma}
	\boldsymbol{\gamma}_j = \mathbf{1} - \left( \mathbf{q}_{j} - \mathbf{p}_{j} \right) \oslash \left( \mathbf{q}_{j} + \mathbf{1} \right), 
\end{align}
where $ \mathbf{q}_{j} = \big[ \max \big\{ \mathcal{B}_1 \big\}, \max \big\{ \mathcal{B}_2 \big\} \big]^\mathrm{T} $, $ \mathbf{p}_{j} = \big[ \min \big\{ \mathcal{B}_1 \big\}, \min \big\{ \mathcal{B}_2 \big\} \big]^\mathrm{T} $, $ \mathcal{B}_k  = \big\{ \left[ \mathbf{B} \right]_{k,1}, \dots, \left[ \mathbf{B} \right]_{k,L} \big\} $.

\textbf{Iterative process:} At each iteration, a new position estimate of the \gls{TN} is obtained according to (\ref{eq:position-update}). The position is updated based on the previous estimate until a stopping criterion is met, which occurs when the improvement in the position estimate between consecutive iterations falls below a specified threshold $ \epsilon $ or when the maximum number of iterations $ J $ is exceeded. The final position estimate is denoted by $ \hat{\mathbf{x}} $.

\textbf{Bounding:}
Elevated noise power can lead to excessively large step sizes in the optimization process, potentially causing the estimated position of the \gls{TN} to drift beyond the feasible region of interest. To ensure the solution remains within a realistic domain, we impose a constraint on the location update, shown below 
\begin{align} \label{eq:bounding}
	\left[ \widetilde{\mathbf{x}}_j \right]_k = \left[ \mathbf{y}_j \right]_k + \left[ \mathbf{e}_j \right]_k,
\end{align}
where 
$ \left[ \mathbf{y}_j \right]_k = \max \left\lbrace \min \left\lbrace  \left[ \widetilde{\mathbf{x}}_j \right]_k, X_k^\mathrm{max} \right\rbrace, X_k^\mathrm{min} \right\rbrace $, 
$ \left[ \mathbf{e}_j \right]_k = z \cdot \mathrm{sgn} \left( \left[ \mathbf{y}_j \right]_k - \left[ \widetilde{\mathbf{x}}_j \right]_k \right) $, and $ z \sim \mathcal{U} \left(\alpha, \beta^2 \right) $. Here, $ \mathbf{e}_j $ is a corrective perturbation vector that projects the location back into the feasible region. The magnitude and direction of the perturbation in each dimension are governed by the random variable $ z $.

\textbf{Analysis:} The parameter $ \boldsymbol{\rho}_j $ is intrinsically linked to $ \boldsymbol{\gamma}_j $ via (\ref{eq:gamma}), forming the core of a self-adaptation mechanism. This relationship allows the algorithm to dynamically respond to the statistical properties of the recent gradients, operating in two distinct regimes, described next.

\emph{Accelerated Convergence}: When the buffered squared gradients exhibit minimal variation, the term $ \mathbf{q}_{j} - \mathbf{p}_{j} $ becomes small. This drives $ \boldsymbol{\rho}_j \rightarrow \mathbf{1} $, which effectively resets the memory of the smoothed term $ \mathbf{c}_j $ in (\ref{eq:smoothed-term}). Consequently, the update prioritizes the most recent gradient information, favoring rapid convergence while potentially compromising stability.

\emph{Enhanced Stability}: Conversely, significant fluctuations in the buffered gradients result in a large value of $ \mathbf{q}_{j} - \mathbf{p}_{j} $. This causes a sharp reduction in $ \boldsymbol{\gamma}_j $, often saturating $ \boldsymbol{\rho}_j $ at its lower bound $ \widetilde{\boldsymbol{\rho}} $. In this state, $ \mathbf{c}_j $ employs a longer history of gradients, creating a more robust averaging process that mitigates the effect of large, potentially disruptive updates. This enhances stability at the expense of a reduced convergence rate.

In essence, \gls{BARProp} dynamically balances convergence speed and stability by adjusting the decay factor based on the recent gradient behavior. By controlling the influence of past gradients on $ \boldsymbol{\gamma}_j $, the algorithm autonomously transitions between these two extremes, providing a robust and adaptive optimization strategy. The complete \gls{BARProp} algorithm is detailed in Algorithm \ref{alg:barprop-algorithm}.

\begin{algorithm}[!h]

	\KwIn{ Find intial feasible position $ \widetilde{\mathbf{x}}_0 $}
	\KwOut{Return the estimated position $ \hat{\boldsymbol{x}} $}
	
 	\While{$ \sim $ {stopping criterion}}
 	{
	Compute the gradient according to (\ref{eq:gradient})
	\\
	Compute the adaptive decay factor according to (\ref{eq:decay-factor})
	\\
	Compute the smoothed term according to (\ref{eq:smoothed-term})
	\\
	Update the TN's position according to (\ref{eq:position-update})
	\\
	Correct the TN's position according to (\ref{eq:bounding}) 
	}
    \caption{Proposed BARProp }
    \label{alg:barprop-algorithm}
\end{algorithm}


\section{{Experimentation and Performance Analysis}}

\label{sec_Simulation}

We evaluate the performance of \gls{BARProp} through both simulations and real-world \gls{RSS} measurements.
For comparison, existing techniques such as the DEOR \cite{caceres2020DEOR}, SDP \cite{jour_Z_Wang_2019}, SOCP \cite{jour_Chang_2018}, 
and \gls{RMSProp} 
\cite{tieleman2012:rmsprop} 
are included. 
Additionally, the ML solution obtained using the Levenberg–Marquardt algorithm, initialized with the true location of the TN and denoted as “ML-true”, is also considered.
We use the \gls{RMSE} as the performance metric, which is defined as 
%
%
%
\begin{IEEEeqnarray}{rCl} \label {eq:RMSE}
	{\rm{RMSE}} = \sqrt {\frac{1}{M}{\sum\limits_{m = 1}^M {{{\left\| {{\mathbf{x}_{m}} - {{\hat{\mathbf{x}}}_{m}}} \right\|}^2_2}} } },
\end{IEEEeqnarray}
where $M$
is the number of Monte Carlo runs,
and $\mathbf{x}_{m}$ and ${\hat{\mathbf{x}}}_{m}$ denote
the real and estimated positions of the TN 
at the $m$-th run, respectively.
%
The \gls{RSS} measurements are generated using the propagation model (\ref{eq_PL}), assuming $P_0=-10$ dBm, $\gamma=3 $,
and $\sigma_n = \left\lbrace 1, 2, 3, 4, 5 \right\rbrace $.


For the \gls{BARProp}, we set the parameters as follows: $ \mu = 0.04 $, $ \widetilde{\boldsymbol{\rho}} = 0.92 $, $ \widetilde{\boldsymbol{\delta}} = 10^{-7} $, $ L = 4 $, $ \epsilon = 0.01 $, $ J = 800 $, $ \alpha = 0 $, and $ \beta = 0.75 $. For the \gls{RMSProp}, the same parameters are used except for the learning rate, which is set to $ \mu = 0.25 $. 
These values were selected based on extensive preliminary experiments to ensure stable convergence and strong performance. For the DEOR method, parameter configurations are directly adopted from~\cite{caceres2020DEOR}.

%

\begin{figure}[!t]
\centering
  \subfigure[]{
  \includegraphics[width=1\linewidth]{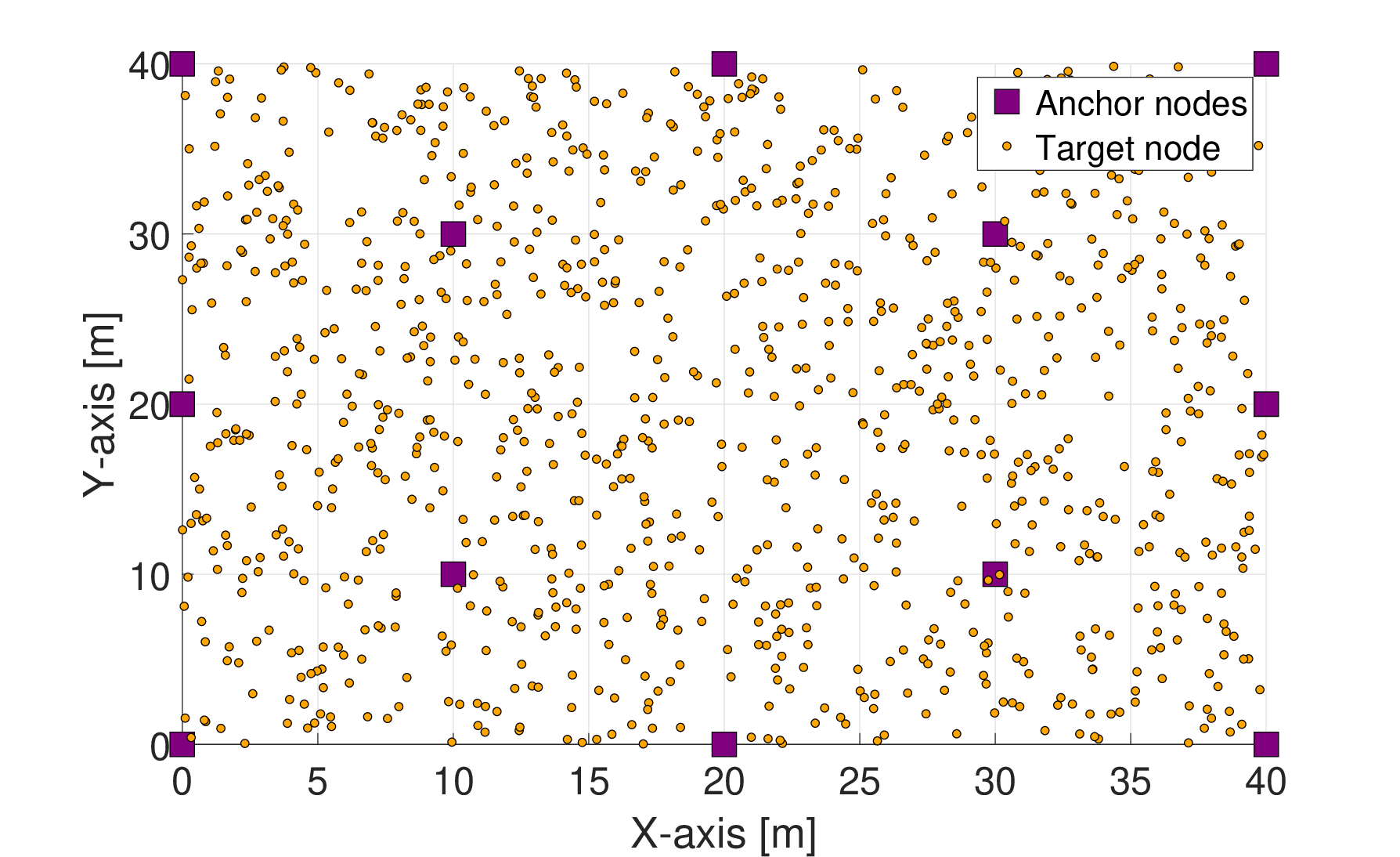}
  }
  \subfigure[]{
  \includegraphics[width=1\linewidth]
  {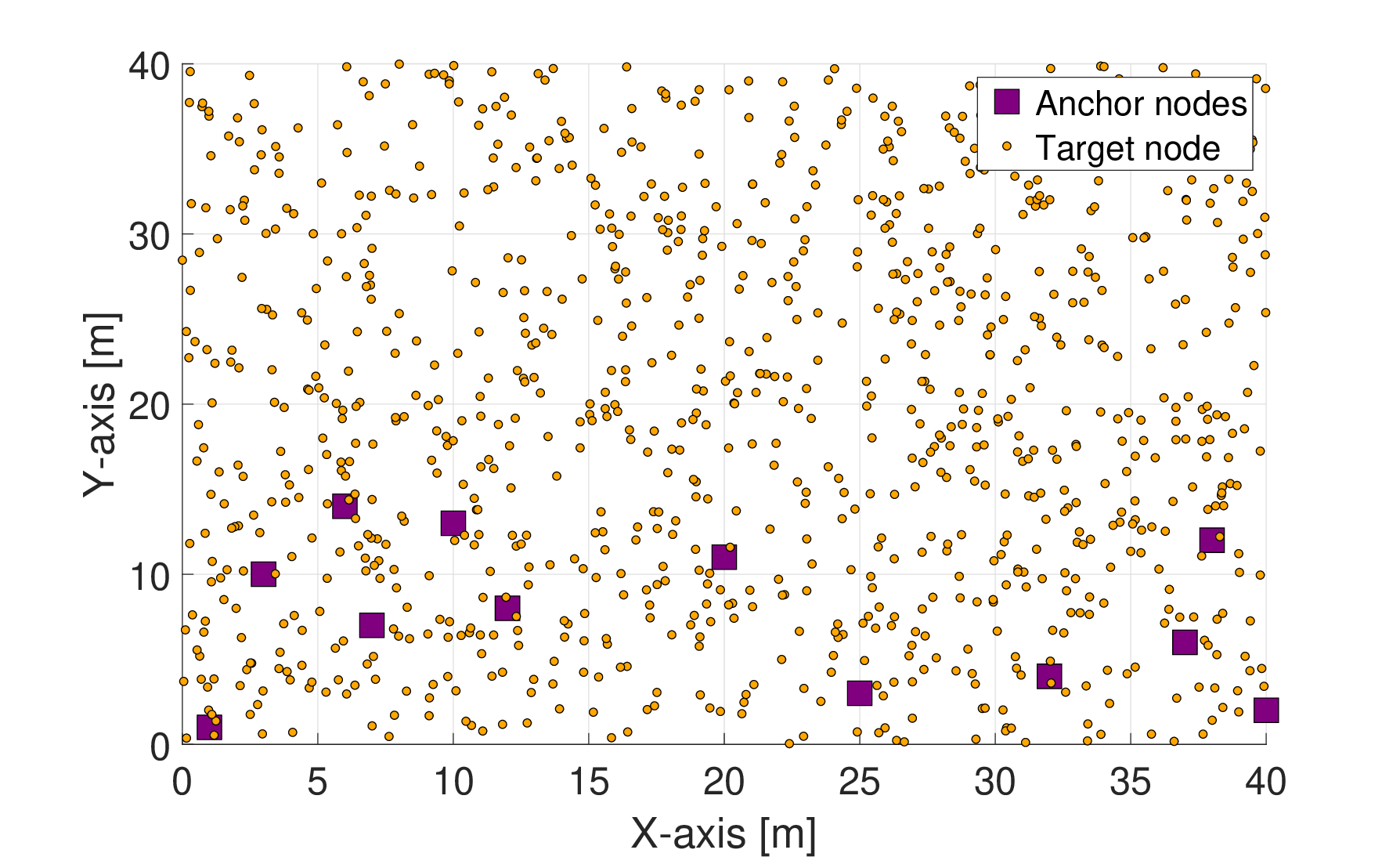}
  }
  \vspace{-3mm} 
  \caption{{Two types of distribution of ANs: (a) homogeneous and (b) non-homogeneous.}}
\label{fig_Scenarios}
\end{figure}

\subsection{{Influence of Anchor Node Distribution and Noise}}
\label{subSec_Influence_ANs_Distri}

The performance of the algorithms is assessed under varying distributions of ANs and levels of log-shadowing noise. We consider an area of interest of $40 \, \text{m} \times 40 \, \text{m}$ and examine both homogeneous and non-homogeneous distributions of ANs.
Here, 
a homogeneous distribution of ANs dictates that ANs are positioned such that their collective convex hull fully encloses the area of interest, thereby guaranteeing coverage for all TNs, as illustrated in Fig. \ref{fig_Scenarios}(a).
In contrast, 
a non-homogeneous distribution of ANs means the ANs are confined to a part of the total area of interest, forming a convex hull that does not cover the whole region, as shown in Fig. \ref{fig_Scenarios}(b).

In practical IoT applications, the AN deployment strategy is critically dependent on the operational environment. For instance, a homogeneous distribution proves effective in scenarios demanding ubiquitous coverage, such as a smart city's environmental monitoring system, where ANs can be uniformly deployed across an urban area to provide consistent data on air quality, noise, and temperature. In contrast, a non-homogeneous distribution is necessitated by environments with significant deployment constraints, such as underground mines. Here, stringent safety regulations and inaccessible areas often restrict AN placement to high-risk zones like active excavation sites or ventilation shafts. 
Consequently, the ANs can only be deployed within a fraction of the total area of interest, which may compromise localization accuracy, as will be demonstrated. Therefore, algorithms robust to such deployment constraints are highly preferable.

Fig. \ref{fig_Scenarios}(a) and Fig. \ref{fig_Scenarios}(b) depict the placement of ANs for the homogeneous and non-homogeneous distribution of ANs considered here. 
To be precise,
in the homogeneous distribution scenario, the ANs are placed at coordinates $(40, 40)$, $(40, 0)$, $(0, 40)$, $(0, 0)$, $(40, 20)$, $(20, 40)$, $(0, 20)$, $(20, 0)$, $(10, 10)$, $(10, 30)$, $(30, 30)$, and $(30, 10)$. This configuration ensures that the TN is always located within the convex hull formed by the ANs, generally resulting in a more favorable geometric arrangement for localization.
In the non-homogeneous distribution of ANs, the ANs are placed at coordinates $(32, 4)$, $(40, 2)$, $(6, 14)$, $(1, 1)$, $(38, 12)$, $(20, 11)$, $(3, 10)$, $(12, 8)$, $(7, 7)$, $(10, 13)$, $(25, 3)$, and $(37, 6)$. This distribution positions the TN outside the convex hull, typically increasing the complexity of localization.


\begin{figure}[!t]
\centering
  \subfigure[]{
  \includegraphics[width=1\linewidth]{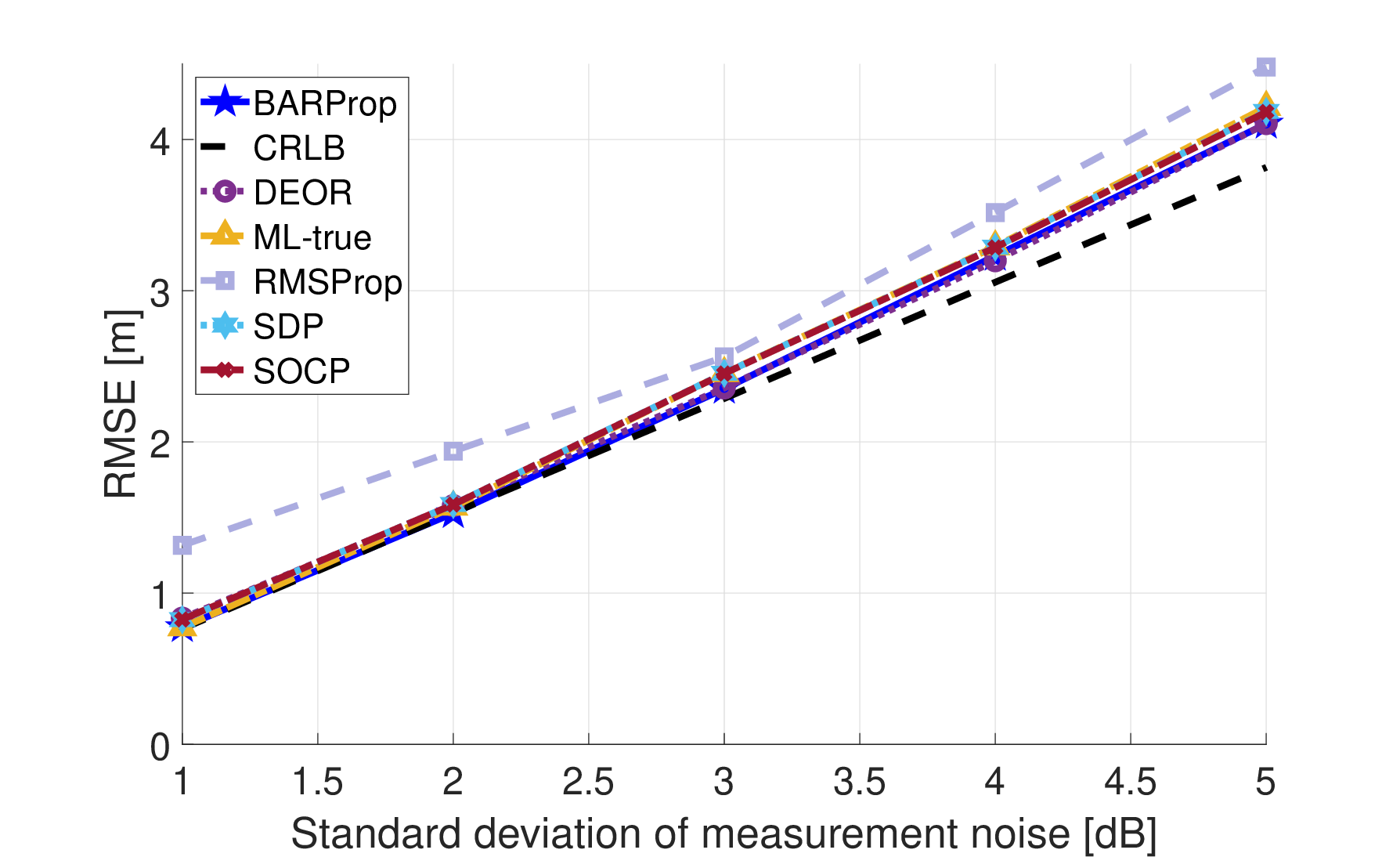}
  }
  \subfigure[]{
  \includegraphics[width=1\linewidth]
  {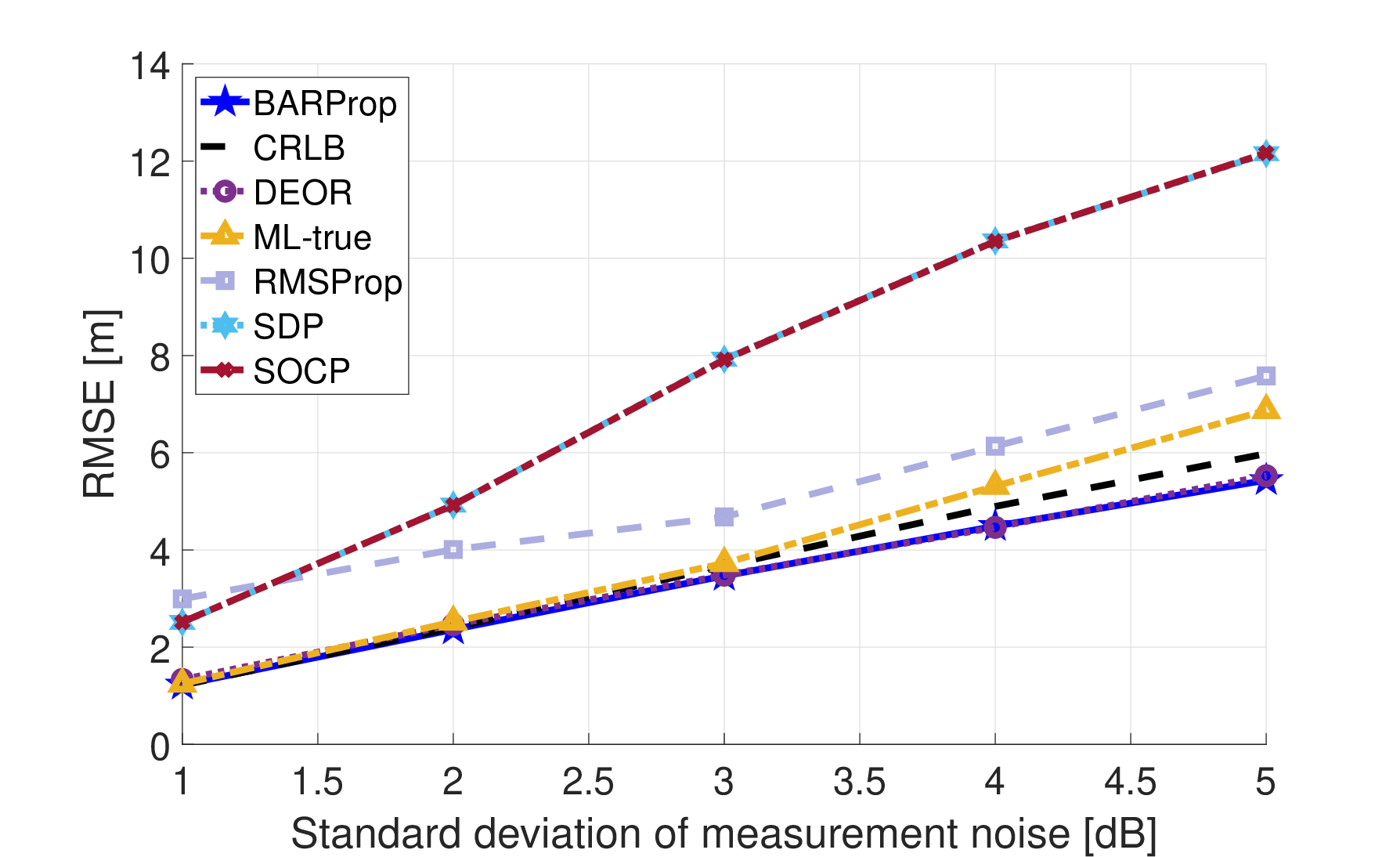}
  }
  \vspace{-3mm} 
  \caption{{RMSE versus the standard deviation of the log-shadowing noise for (a) homogenous and (b) non-homogeneous distribution of ANs.}}
  \vspace{-2mm} 
\label{fig_NRMSE_vs_Noise_Homo}
\end{figure}

Fig. \ref{fig_NRMSE_vs_Noise_Homo} shows the localization accuracy of the algorithms evaluated under both  homogeneous and non-homogeneous distribution of ANs
as the standard deviation of the log-shadowing noise varies from 1 to 5 dB.
%
It is observed that
the performance of the algorithms
deteriorates
as the standard deviation increases.
However,
such trend is less prominent for the BARProp and DEOR, specially in the scenario with non-homogeneous distribution of ANs.
Additionally,
the distribution of ANs considerably affects the localization accuracy of all the algorithms.
In the scenario with homogeneous distribution of ANs, all the algorithms provide similar localization accuracy close to the CRLB.
In contrast, in the scenario with non-homogeneous distribution of ANs, the accuracy of all the algorithms deviates further from the CRLB as the standard deviation of the log-shadowing noise increases.
Note that the proposed BARProp algorithm together with the DEOR exibit the least sensitivity to these changes unlike the SDP, SOCP, and ML-true algorithms.
For instance, when transitioning from homogeneous to a non-homogeneous distribution of ANs at $5$ dB of the standard deviation of the log-shadowing noise, the localization accuracy of the BARProp, DEOR, ML-true, RMSProp, SDP, and SOCP are reduced by 
$23.95\%$,
$24.56\%$,
$38.37\%$,
$39.03\%$,
$64.22\%$, and
$64.23\%$, 
respectively.
It should be noted that the CRLB applies only to unbiased estimators. Since both BARProp and DEOR exhibit a slight bias, their RMSE can occasionally fall below the CRLB. Similar observations have been reported in prior works (e.g., \cite{jour_Ouyang, Tomic_RSS_coop_noncoop_loc_2015}).


\subsection{{Influence of the Number of Anchor Nodes}}
\label{subSec_Influence_NumbANs}

In Section \ref{subSec_Influence_ANs_Distri},
we evaluated the performance of the proposed algorithm 
when $N$ was fixed within the area of interest.
We now extend this analysis to investigate the influence of varying a $ N $ on localization accuracy, considering 
$ N = \left\lbrace 10,    14,    18,    22,    26,   30 \right\rbrace $. 
For each number of ANs, the ANs are deployed randomly in the area of interest, and then, at each of the $1000$ Monte Carlo runs, the location of the TN is also deployed randomly.

%

\begin{figure}[!t]
	\centering
	\includegraphics[width=1\linewidth]
        {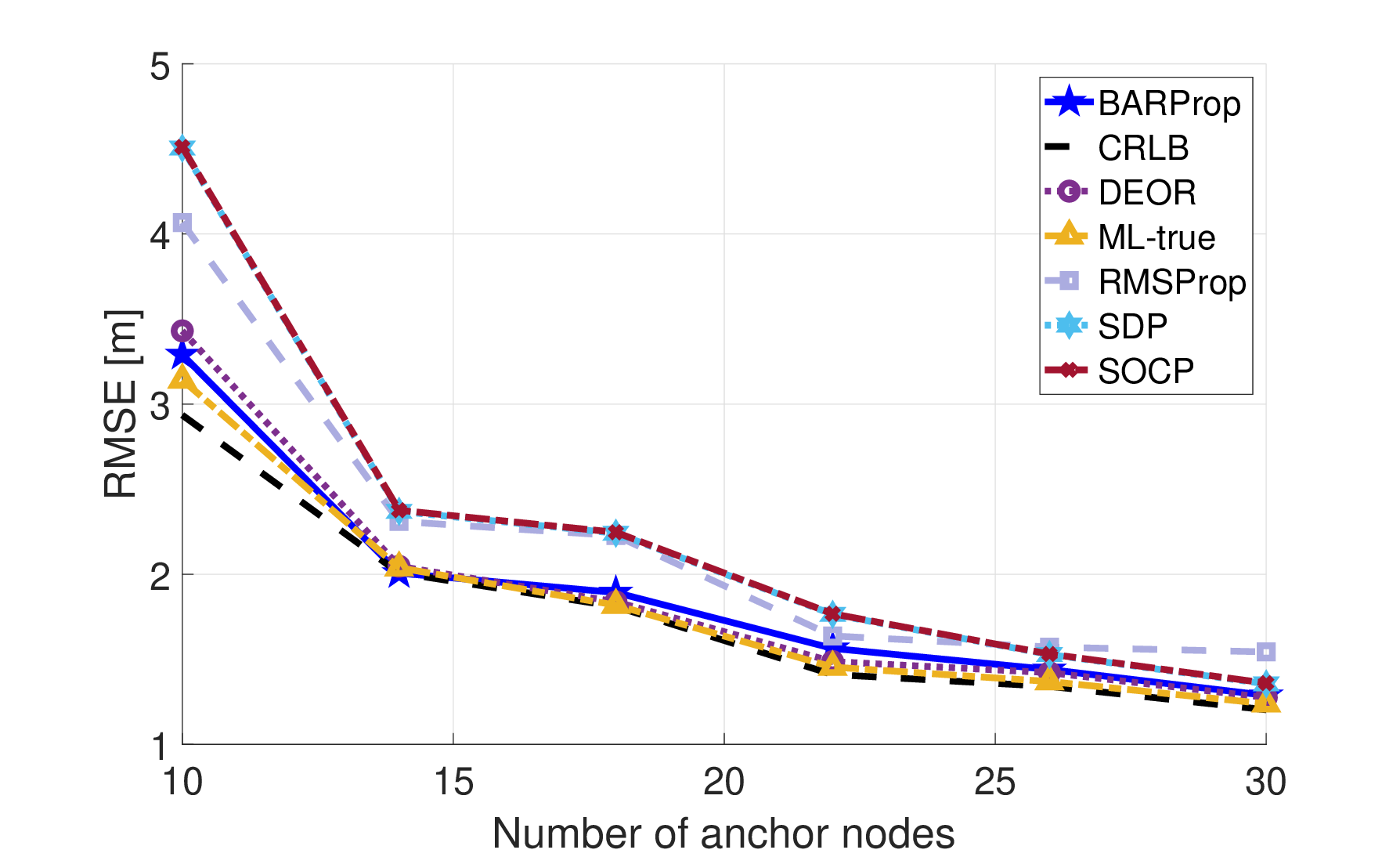}
	\caption{{RMSE versus the number of ANs.}}
	\label{fig_RMSE_vs_Numb_ANs}
 
\end{figure}

Fig. \ref{fig_RMSE_vs_Numb_ANs} presents RMSE values as a function of the number of ANs, with the standard deviation of the log-shadowing noise set to $\sigma=3$ dB. As anticipated, localization accuracy improves with an increasing number of ANs due to increased availability of RSS measurements. Among the evaluated algorithms, the proposed BARProp algorithm, along with the DEOR and ML-True, exhibit superior localization accuracy. 
The BARProp algorithms 
demonstrate a considerable advantage over the other considered algorithms, particularly outperforming the RMSProp, SDP, and SOCP, across the full spectrum of AN configurations.
To be specific, when the number of ANs is $18$, the BARProp algorithm improves the localization accuracy by
$15.16\%$,
$15.91\%$,
and
$16.28\%$
with respect to 
RMSProp, SDP, and SOCP, respectively.

Although the proposed BARProp algorithm shows only marginal improvements in some cases and slightly lower localization accuracy compared to DEOR and ML-true, its notable advantage lies in superior computational efficiency. 
To substantiate this, 
the average running time for all algorithms was measured under the same settings and parameters as those used in Fig. \ref{fig_RMSE_vs_Numb_ANs}, with the results summarized in Table \ref{tbl:avg_time_ANrange}. 
It is observed that the proposed BARProp algorithm demonstrates very high computational efficiency among the tested methods, second only to RMSProp. Specifically, the BARProp is approximately $5$, $4$, $339$, and $444$ times faster than DEOR, ML-true, SDP, and SOCP, respectively. 
These empirical findings will be further corroborated by a theoretical computational complexity analysis of the algorithms in the subsequent section.

\begin{table}[!t]
	\centering
	\caption{Average running time in milliseconds when the number of anchor nodes varies from 10 to 30.}
	\begin{tabular}{lc}
		\toprule
		{Algorithm} & Time [ms] \\
		\midrule\midrule
		{BARProp}  & 0.93 \\
		\midrule
		{DEOR \cite{caceres2020DEOR}}     & {4.48} \\
		\midrule
		{ML-true}  & 3.59 \\
		\midrule
		{RMSProp \cite{tieleman2012:rmsprop}}  & 0.78 \\
        \midrule
		{SDP \cite{jour_Z_Wang_2019}}      & 315.92\\
		\midrule
		{SOCP \cite{jour_Chang_2018}}     & 413.92 \\
		\bottomrule
	\end{tabular}
	\label{tbl:avg_time_ANrange}
\end{table}

\subsection{Computational Complexity}
\label{sec_Complexity}
%
%

The computational complexity of an algorithm is a critical factor that complements localization accuracy. There is often a trade-off between achieving high accuracy and maintaining low computational demands. Algorithms with high complexity may be impractical for real-time applications, whereas those with lower complexity are better suited for such scenarios.
This is particularly pertinent in IoT applications, where sensor nodes are typically constrained by limited computational power and battery life.
This section therefore evaluates and compares the computational complexity of the considered algorithms.

The \textit{Big-O} notation is employed to characterize the theoretical complexity of algorithms. In this analysis, minor terms are disregarded, focusing instead on the dominant terms. 
In general, 
the primary term shared by all algorithms is the number $N$ of ANs. 
For \gls{BARProp}, the computational complexity additionally depends on the number of initialization points $ U $ and the number of iterations $ J $. It is important to note, however, that this represents a worst-case bound. In practice, \gls{BARProp} often converges to a solution before reaching $ J $ iterations once the prescribed error threshold $ \epsilon $ is satisfied. \gls{RMSProp} has similar complexity as \gls{BARProp} and can also converge earlier if the prescribed error threshold is achieved.
For the DEOR case, the key parameters influencing further its computational complexity are the maximum number of generations $G$ and size of the population $K$.
In contrast, the complexity of ML-true, SDP, and SOCP are of a higher order, specifically exceeding $ N^3 $.
Table \ref{tab_Comp_Complexity} presents the theoretical computational complexity of all the algorithms reviewed in the preceding section.
Notably, the theoretical complexity of the proposed \gls{BARProp} algorithm and DEOR scales linearly with the number $ N $ of ANs. 
 Consequently, it is anticipated that the proposed algorithm \gls{BARProp} will exhibit reduced computational times, particularly in scenarios involving a large number of anchor nodes. This expectation will be substantiated in the subsequent analysis.

\begin{table}[!t]
 \centering
 \caption{{Computational complexity and required memory allocation of benchmarked algorithms}} 
  \begin{tabular}
  {l c c}
      \toprule
  Algorithm & Complexity & Memory [Bytes] \\
    \midrule
    \midrule
    BARProp & {$O\left (U + N J \right)$} & $  8L + 9 $ \\
    \midrule
    DEOR\cite{caceres2020DEOR}  & $O(G K N  + K\log (2K))$ & $ 24(K+2) $ \\
    \midrule
   {ML-true}  &  {$O\left (N^3 \right)$} & $  16(N+5) $ \\
    \midrule
    RMSProp \cite{tieleman2012:rmsprop} & {$O\left( N J \right)$} & $  10 $ \\
    \midrule
  SDP \cite{jour_Z_Wang_2019} & $O\left (N^4 \right)$ & $ \approxprop 8 N^3 $   \\
      \midrule
  SOCP \cite{jour_Chang_2018} & $O\left( N^3 \right)$ & $ \approxprop 4 N^2 $  \\
 \bottomrule
  \end{tabular}%
  \label{tab_Comp_Complexity}%
\end{table}%

\begin{figure}[!t]
	\centering
	\includegraphics[width=1\linewidth]
	{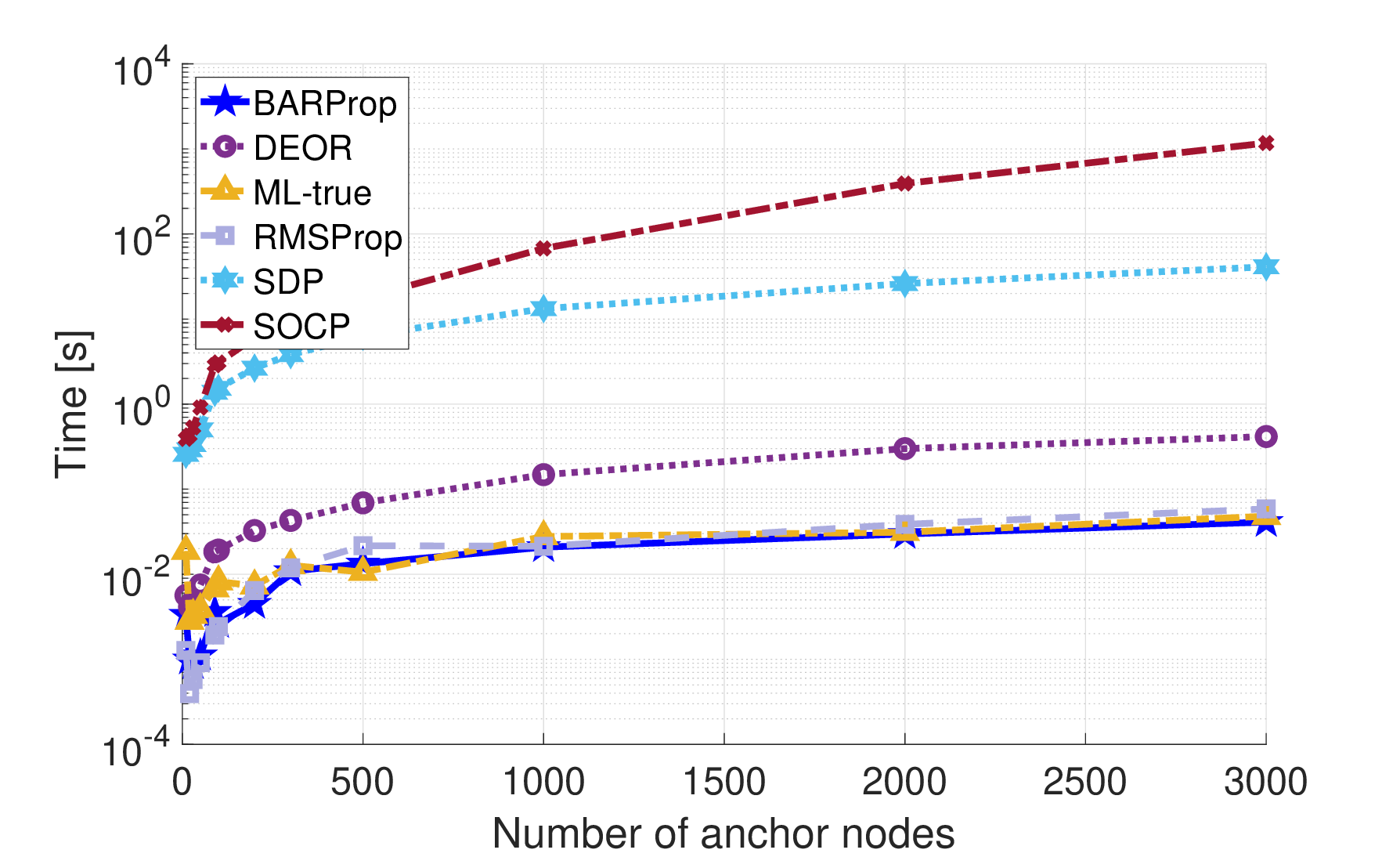}
	\caption{Average running time versus number of
		target nodes (CPU: Intel (R) Core (TM) i5-14400 2.50 GHz. RAM: 16.0 GB.).}
	\label{fig_SimuTimevsANs}
\end{figure}

Using the same parameter values as those in Fig.~\ref{fig_RMSE_vs_Numb_ANs}, we obtained measurements of the actual computational time. Fig.~\ref{fig_SimuTimevsANs} depicts the average computational time for all algorithms on a logarithmic scale.
This figure illustrates that \gls{BARProp} is the least computationally demanding compared to all other evaluated algorithms. 
This observation aligns with the theoretical complexity outlined in Table~\ref{tab_Comp_Complexity}.

In addition to computational complexity, memory requirements are a critical consideration, particularly given the limitations of typical IoT devices. Therefore, Table~\ref{tab_Comp_Complexity} also reports the memory usage of all approaches. It is important to note that these values refer solely to the memory required for algorithmic computations and do not account for storing the RSS measurements, which is common to all methods and amounts to $ 2N $ bytes. For \gls{BARProp}, the memory is primarily determined by the \gls{FIFO} buffer and additional parameters required for gradient computation. In bytes, the memory footprint is approximately $ 8L + 9 $. \gls{RMSProp} requires even less memory, as it does not maintain a buffer of past gradients. Memory requirements for the remaining approaches are considerably higher. Comparing \gls{BARProp} with DEOR, its closest competitor in terms of accuracy, we find that \gls{BARProp} requires only about $14.2\%$ of DEOR’s memory under our simulation settings ($ L = 4 $) and with $ K = 10 $ for DEOR, as specified in~\cite{caceres2020DEOR}.

%


\subsection{Real Indoor Experiment}

\begin{figure}[!t]
	\centering
	\includegraphics[width=0.9\linewidth]
	{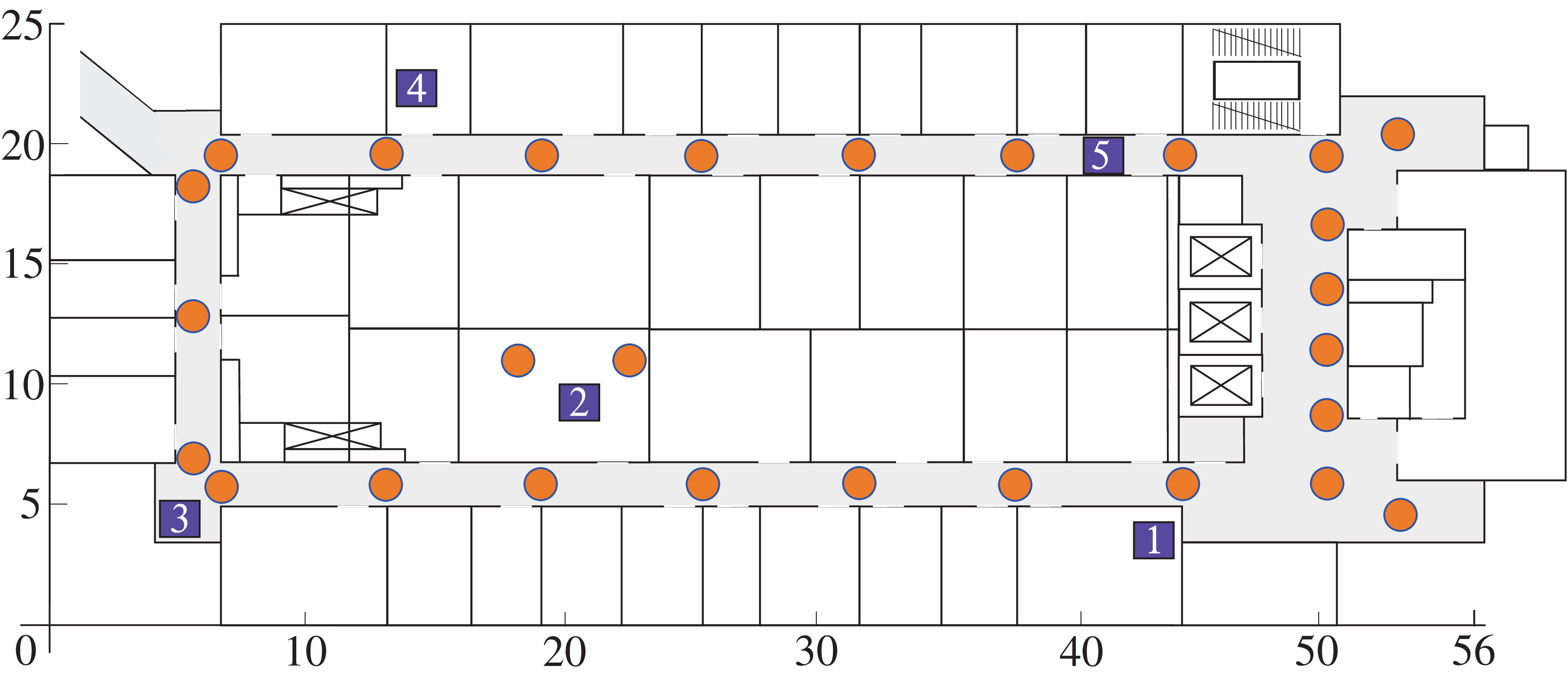}
	\caption{Real indoor environment set-up with 5 ANs (Violet squares) and 27 locations of a TN (Orange circles)}
	\label{fig_Real_indoor_experiment_set_up}
\end{figure}

The RSS measurements were acquired in a real indoor environment covering an area of \(56\,\mathrm{m} \times 25\,\mathrm{m}\) \cite{niculescu2004vor}, as depicted in Fig.~\ref{fig_Real_indoor_experiment_set_up}.  
The experimental setup involved five ANs and 27 predefined positions of a single TN. In the figure, violet squares denote the AN locations, while the orange circles indicate the TN positions.  
At each AN, $1000$ RSS measurements were collected from each of the $27$ TN positions. For each localization trial, a single RSS sample from each AN was selected and used to estimate the TN’s position. This process was repeated $1000$ times to ensure statistical reliability and robustness of the results.

\begin{figure}[!t]
	\centering
	\includegraphics[width=1\linewidth]
	{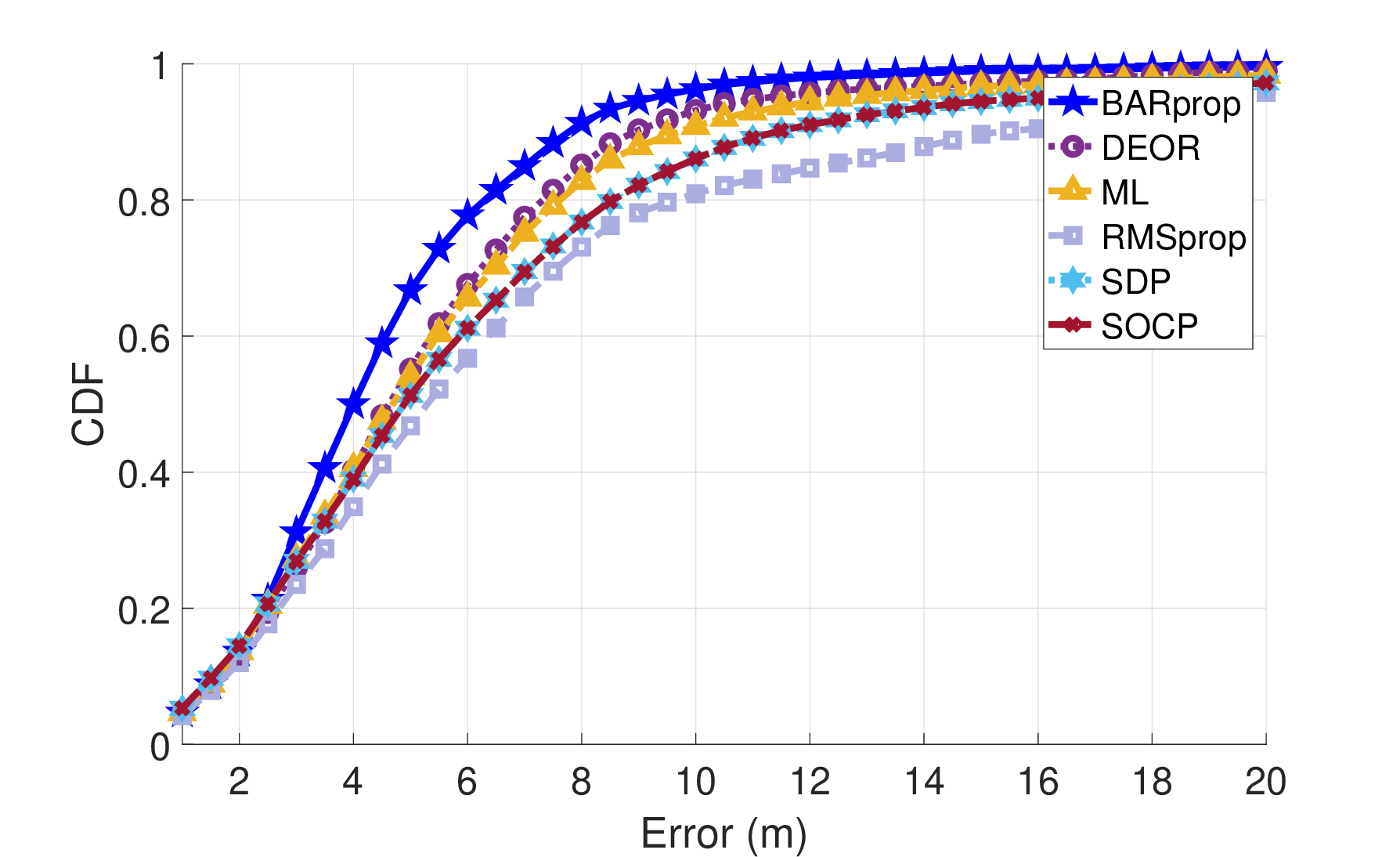}
	\caption{CDF of localization error for various localization algorithms using real RSS measurements}
	\label{fig_CDF_Real_indoor_experiment}
\end{figure}

Fig.~\ref{fig_CDF_Real_indoor_experiment} shows the cumulative distribution function (CDF) of the localization error, expressed as \( \left\|\hat{x}_m - x_m\right\|_2 \). The results show that the proposed algorithm provides superior localization performance compared to the benchmark methods. Specifically, the \gls{BARProp} achieves a localization error of less than or equal to 6.5\,m in 82\% of the trials. In contrast, the corresponding probabilities for the DEOR, ML, RMSProp, SDP, and SOCP methods are below 72\%, 70\%, 70\%, and 61\%, respectively, thereby demonstrating the effectiveness of the proposed method in enhancing localization accuracy.

 \begin{table}[!t]
	\centering
	\caption{Average Running Time in Milliseconds of Several Algorithms Measured in a Real Indoor Environment}
	\begin{tabular}{lc}
		\toprule
		{Algorithm} & {Time [ms]} \\
		\midrule\midrule
		{BARProp}  & {0.66} \\
		\midrule
		{DEOR \cite{caceres2020DEOR}}     & {3.23} \\
		\midrule
		{ML-true}  & {2.70} \\
		\midrule
		{RMSProp \cite{tieleman2012:rmsprop}}  & {0.41} \\
        \midrule
		{SDP \cite{jour_Z_Wang_2019}}      & {463.90}\\
		\midrule
		{SOCP \cite{jour_Chang_2018}}     & {524.10} \\
		\bottomrule
	\end{tabular}
	\label{tbl:avg_time_RealExperiment}
\end{table}

Additionally, Table \ref{tbl:avg_time_RealExperiment} presents the average running time for all considered algorithms obtained from real indoor experiments. Consistent with the results discussed in previous sections, the proposed BARProp algorithm demonstrates the highest computational efficiency among the tested methods, second only to RMSProp. Specifically, for this particular experimental setup, BARProp is approximately 
$4.91$, $4.09$, $703.89$, and $795.23$ times faster than 
DEOR, ML-true, SDP, and SOCP, respectively.
It is worth mentioning that although RMSProp exhibits superior computational efficiency, BARProp consistently provides substantially higher localization accuracy.

\section{Discussion and Future Work}

The proposed BARProp algorithm has been developed and evaluated for IoT localization using RSS measurements under varying anchor node distributions. In future work, we aim to extend this framework to three-dimensional localization to address challenges such as altitude-dependent fading, multipath propagation, and irregular anchor placement. This extension will broaden its applicability to more complex environments, including aerial networks, underwater sensor systems, and multilevel indoor infrastructures, where geometry and noise conditions are considerably more demanding.

It is noteworthy that RSS-based localization techniques offer the advantages of lightweight implementation and low power consumption, which makes them well-suited for resource-constrained IoT devices. However, relying solely on RSS limits robustness in multipath-rich or obstructed environments. In future work, we aim to incorporate multi-modal measurements such as TOA, AOA, and CSI to improve accuracy and resilience. However, adopting these modalities presents important challenges. For instance, TOA requires precise time synchronization across devices, which is costly and difficult to maintain in large-scale IoT deployments. Similarly, AOA demands specialized antenna arrays and calibration procedures, increasing hardware complexity and energy consumption. Our goal will be to balance the efficiency of RSS-based methods with the richer information provided by these modalities, establishing a trade-off between lightweight design and advanced localization accuracy to ensure practicality across diverse IoT scenarios.

The complexity analysis shows that BARProp scales linearly with the number of ANs, with overall complexity $O(U+NJ)$, making it both computationally efficient and memory-friendly for large-scale, energy-constrained WSNs. Looking ahead, a promising research direction is the development of an online, distributed implementation of BARProp. Such an approach would eliminate the reliance on centralized processing and significantly reduce communication overhead. We believe that this would enable even greater scalability and adaptability in real-time IoT deployments.

Although the adaptive decay factor in BARProp provides a balance between stability and convergence speed, further improvements could be achieved through adaptive strategies for buffer length $L$ and learning rate $\mu$ to better adjust under dynamically changing environments. Another promising direction is the development of energy-aware variants of BARProp to extend battery life in resource-constrained IoT nodes without compromising accuracy. Moreover, since many IoT applications (e.g., asset tracking, robotics, and autonomous navigation) require reliable tracking of mobile nodes, future work should focus on extending BARProp to dynamic target tracking and predictive modeling, thereby broadening its applicability to real-world, time-varying scenarios.

\section{Conclusion}
\label{sec_Conclusion}

This work investigated \gls{RSS}-based localization and proposed \gls{BARProp}, a novel learning-inspired optimization algorithm designed for high accuracy, fast convergence, and low memory footprint. Extensive evaluations demonstrated that \gls{BARProp} surpasses state-of-the-art methods in localization accuracy while achieving a fourfold speedup and significantly reduced memory consumption. These attributes make \gls{BARProp} exceptionally suitable for resource-constrained devices prevalent in \gls{WSN} and IoT applications. The algorithm was rigorously evaluated against convex optimization-based methods (e.g., SDR, SOCP) and evolutionary approaches (e.g., DEOR) using both synthetic and real-world \gls{RSS} data. In all cases, \gls{BARProp} exhibited substantial performance superiority.




\bibliographystyle{IEEEtran}
\bibliography{RefBibioReview.bib}

\begin{IEEEbiography}[{\includegraphics[width=1in,height=1.25in,clip,keepaspectratio]{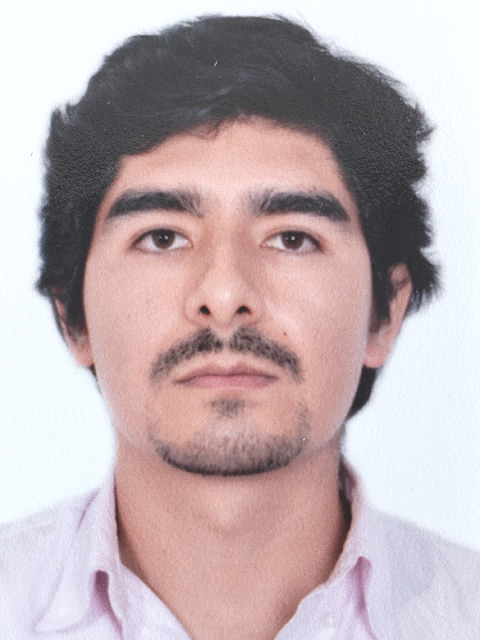}}]{Luis F. Abanto-Leon} received the M.Sc. degree in communications engineering from Tohoku University, Japan, in 2015. From 2016 to 2018, he was a Researcher with Eindhoven University of Technology (TU/e) and NXP Semiconductors, Eindhoven, The Netherlands. He received the Ph.D. degree in computer science from Technische Universität Darmstadt, Germany, in 2023. Since March 2024, he has been a Postdoctoral Researcher with the Faculty of Electrical Engineering and Information Technology, Ruhr-Universität Bochum, Germany. His research expertise include optimization theory, signal processing, and machine learning, with a focus on algorithm design for radio resource management in 5G/6G wireless networks.
\end{IEEEbiography}

\begin{IEEEbiography}[{\includegraphics[width=1in,height=1.25in,clip,keepaspectratio]{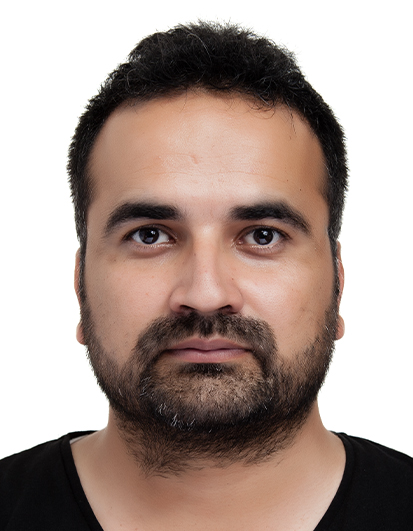}}]{Muhammad Salman} received the B.Sc. degree in Electronic Engineering from the Balochistan University of Information Technology, Engineering and Management Sciences (BUITEMS), Quetta, Pakistan, in 2010, the M.Sc. degree in Electronic Engineering from Politecnico di Torino, Turin, Italy, in 2014, and the Ph.D. degree in Electrical and Computer Engineering from Inha University, Incheon, South Korea, in 2023. He is currently working as an Assistant Professor at the National University of Sciences and Technology (NUST), Pakistan.
Previously, he served as a Postdoctoral Researcher at Korea Institute of Energy Technology (KENTECH), South Korea, from February 2023 to February 2024. Earlier, he worked as a Lecturer in the Electrical and Computer Engineering Department at Effat University, Saudi Arabia, from October 2014 to June 2019. His research interests include computer and wireless networks, as well as contactless sensing.
\end{IEEEbiography}

\begin{IEEEbiography}[{\includegraphics[width=1in,height=1.25in,clip,keepaspectratio]{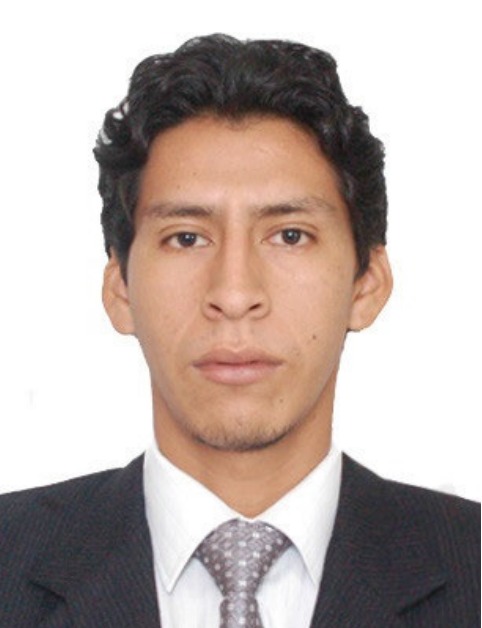}}]{Lismer Andres Caceres Najarro} received the B.Sc.
degree from the Peruvian University of Applied
Sciences, Lima, Peru, in 2010, the M.S. degree from
Kyungsung University, Busan, Republic of Korea, in
2016, and the Ph.D. degree in electrical engineering and computer science from Gwangju Institute
of Science and Technology, Gwangju, Republic of
Korea, in 2021.
He is currently an Assistant Professor at Chosun University with the Department of Computer Engineering, Chosun University, Gwangju, South Korea.
Before joining Chosun University he was a Research Professor with the
Korea Institute of Energy Technology, Naju, South
Korea. His research interests include target localization in wireless sensor networks, evolutionary algorithms, machine learning,
multiagent systems, smart grids, and smart healthcare.
Prof. Caceres Najarro was awarded the Graña y Montero Peruvian
Engineering Research Award (fourth edition).
\end{IEEEbiography}

\end{document}